\documentclass[journal]{IEEEtran}
\usepackage{cite}
\usepackage{amsmath}
\usepackage{epsfig}
\usepackage{psfrag}
\usepackage{epstopdf}

\begin{document}

\title{Dissipative Soliton Mediated Radiations in Active Silicon-Based Waveguides}

\author{Ambaresh~Sahoo and~Samudra~Roy
\thanks{Ambaresh Sahoo and Samudra Roy are with the
Departament of Physics, Indian Institute of Technology Kharagpur, West Bengal 721302, India
(e-mail: ambaresh@phy.iitkgp.ernet.in; samudra.roy@phy.iitkgp.ernet.in).}

}

\maketitle

\begin{abstract}

The Ginzburg-Landau (GL) equation  is in general not integrable by the inverse scattering method and support solitary-wave solution, called dissipative soliton (DS). We numerically demonstrate that, a  DS can radiate dispersive waves (DWs) in presence of third-order dispersion (TOD). We propose a silicon-based active waveguide that excites stable DSs.  Energy can be transferred from these stable DS to linear DWs when a resonance condition is achieved.  The dynamics of the DS is governed by the complex GL equation which we solve numerically for different operational parameters. Numerical solution of the perturbed GL equation exhibits multiple radiations, when the stable DS is allowed to propagate through a large distance. We theoretically derive a special phase-matching relation that can predict the frequencies of these multiple radiations, which are found numerically. In our theoretical and numerical calculations we include the role of free carriers which appear inside semiconductor waveguides as a consequence of two-photon absorption (TPA). We demonstrate that apart from TOD, TPA and gain dispersion are two additional parameters that can control the radiation emitted by DS. The DS-mediated radiation is different in nature and demands an intuitive understanding. In this work we try to provide some insights of this fascinating radiation phenomenon by elaborate analytical and numerical calculations. 

\end{abstract}

\begin{IEEEkeywords}
Dissipative soliton, Phase-matched radiation, Ginzburg-Landau equation, Linear dispersive wave, Active Si waveguide.
\end{IEEEkeywords}

\IEEEpeerreviewmaketitle

\section{Introduction}
\IEEEPARstart{N}{onlinear} silicon photonics, as a field, is outspreaded
and is of use due to novel nonlinear properties of silicon (Si) and its compound elements \cite{Leuthold}. Silicon-on-insulator (SOI) technology \cite{Jalali} has grown up very rapidly because of the ability of Si to amplify, generate, process and sense signals \cite{Miller,Krishnamoorthy,Cutolo}. High contrast in refractive index between core and cladding, and large intrinsic nonlinearity (100 times larger than bulk silica glass) of Si allow tight confinement of an optical mode. This tight light confinement ensures novel nonlinear pulse dynamics (like soliton formation, supercontinuum generation etc.) for low pump power. In Si-based waveguide, the two-photon absorption (TPA) is found to be the leading loss mechanism in wavelength limit 0.8 $\mu$m$<\lambda_0<$2.2 $\mu$m \cite{Liang,Yin}. In TPA process, the semiconductor absorbs two pumped photons simultaneously when the total energy of the photons exceeds the band-gap energy level. This nonlinear absorption process leads to the generations of electron-hole pairs or free carriers (FCs) inside the waveguide.  FCs limit device performance by introduceing loss and also change the refractive index \cite{Dieter,Tomita}. The loss due to the multiphoton absorption is compensated by adopting  organic Si-based hybrid slot waveguides  \cite{Agazzi} which can support some stable  solitary-wave structure called dissipative solitons (DSs). DSs are self-organized stationary structures in open nonlinear systems far from equilibrium where the system is nonconservative and energy can flow in to it \cite{Akhmediev}. 

In this  work, we try to investigate the formation and evolution of DSs inside an active Si-based waveguide where third-order dispersion (TOD) is nonvanishing. In contrast to the Schr\"{o}dinger Kerr soliton, which is generated in a conservative system, when group-velocity dispersion (GVD) is counterbalanced by nonlinearity \cite{Hasegawa}, \cite{Kivshar}, DSs are generated in a dissipative system where, external energy flow is necessary. In order to get a stable DS, it is essential to compensate the loss by supplying some energy as gain into the system \cite{Akhmediev}.  The additional balance of gain and loss results in solutions which are fixed (but not a family of solution) in parameter space\cite{Boardman}. Unlike bright Kerr solitons, the DSs are found both in anomalous dispersion (AD) \cite{Xu,Ma} and normal dispersion (ND) regime \cite{D-Li,Tan}. The spectral characteristics of DS differ significantly in ND, where we observe a flat top spectra \cite{Renninger}. In this work, however, we concentrate more on the DS which is excited in the AD regime.  DS solution, that arises from a nonintegrable system, is very sensitive to the parameters like gain, loss, dispersion, nonlinearity etc. The soliton solution even disappears if the source of energy (here, gain) is removed or if the parameters of the system are chosen beyond the tolerance range. In real systems, the propagating pulse should experience the actual dispersion profile of the waveguide which is nothing but the spectral variation of GVD parameter. The Taylor series expansion of GVD parameter ($\beta_2(\omega)$) leads to TOD as the first higher order correction. It is interesting to investigate, how TOD perturbs the dynamics of a stable DS. In our study, we find phase-matched radiations are emerged from the DS when it is perturbed by TOD. In a previous work, a similer problem was addressed where TOD-induced tempotal shift of DS has been studied analytically by using perturbation analysis \cite{Malomed}. It was shown that, a complex group velocity can be induced by TOD. However, the perturbation analysis has some limitations and it fails completely to predict the phenomenon of phase-matched or resonant radiation. In presence of higher order dispersion, a phase-matching (PM) condition is satisfied for a real frequency ($\omega$), at which, the Kerr soliton radiates \cite{GPA}. This energy transfer leads to a linear wave in ND regime called dispersive wave (DW). The PM condition is achieved by equating the phase of soliton and the DW \cite{GPA,Roy}. It is interesting to study how this resonance condition is modified in case of DS containing an additional phase term that depends on the damping parameters (like TPA loss and gain dispersion). Infact we demonstrate that, apart from TOD, TPA and gain dispersion coefficient are two additional parameters that influence the DS-mediated radiations. This is an unique feature which was never explored before. Further more, in Si-based semiconductor waveguides electron-hole pairs are generated due to multi-photon absorption when it is exposed to the optical fields. These FCs can influence the soliton dynamics significantly by changing the local refractive index of the medium and may also affect DW generation. In our theoretical analysis we include the effect of TOD and FC as perturbations and investigate their influence on stable DSs and DWs. For this work, we consider the Si-based waveguide that is embedded in amorphous active Er-doped  aluminium oxide $(Al_2 O_3:Er^+)$ \cite{Pintus}. Finally  a modified PM relation is established for DSs which accurately predicts the spectral locations of the DWs.

We organize our paper as follows; in Sec.~\ref{model}, we propose and model a physically acceptable active Si-hybrid waveguide, that can excite DSs, and study the dispersion and nonliear properties of this waveguide.  To study the dynamics of DS, in Sec.~\ref{theory}, we introduce the perturbed GL equation, which, is in fact a two (1,1) dimensional nonlinear Schr\"{o}dinger equation having growing and damping terms. We solve the governing equation numerically to show the existence of stable DS. Finally, in Sec.~\ref{dw}, we study how DWs are emerged from a perturbed DS. The role of FCs on DW are investigated thoroughly. We further explore the oscillating nature of DS that radiates multiple DWs under the perturbation of TOD.  The spectral location of these multiple radiations are predicted by deriving a modified PM equation.

\section{Waveguide description}
\label{model}

We design our proposed hybrid Si-based waveguide along the $[\overline{1}10]$ direction. By fabricating the waveguide in this preferred direction, we can diminish the Raman scattering for quasi-TM modes \cite{Lin}. We try to remove the Raman effect to restrict the chaotic dynamics of DS. For this nano-structured waveguide, we obtain the dispersion profile in the wavelength range 1.2 to 1.8 $\mu$m. At the operating wavelength ($\lambda_0$) we calculate GVD coefficient, TOD coefficient and the effective area for guided mode as,  $\beta_2 \approx -0.125 \ ps^2 m^{-1}$, $\beta_3 \approx 0.437\times 10^{-2} \ ps^3 m^{-1}$ and $A_{eff} \approx 0.295 \ \mu m^2$ respectively. The choice of 
$Al_2 O_3:Er^+$ as an active medium offers some advantage since the refractive index is enhanced through $Er+$ ion. Also, due to the amorphous nature of $Al_2 O_3:Er^+$ the emission spectrum becomes broad \cite{Pintus}.

\begin{figure}[h!]
\begin{center}
 \epsfig{file=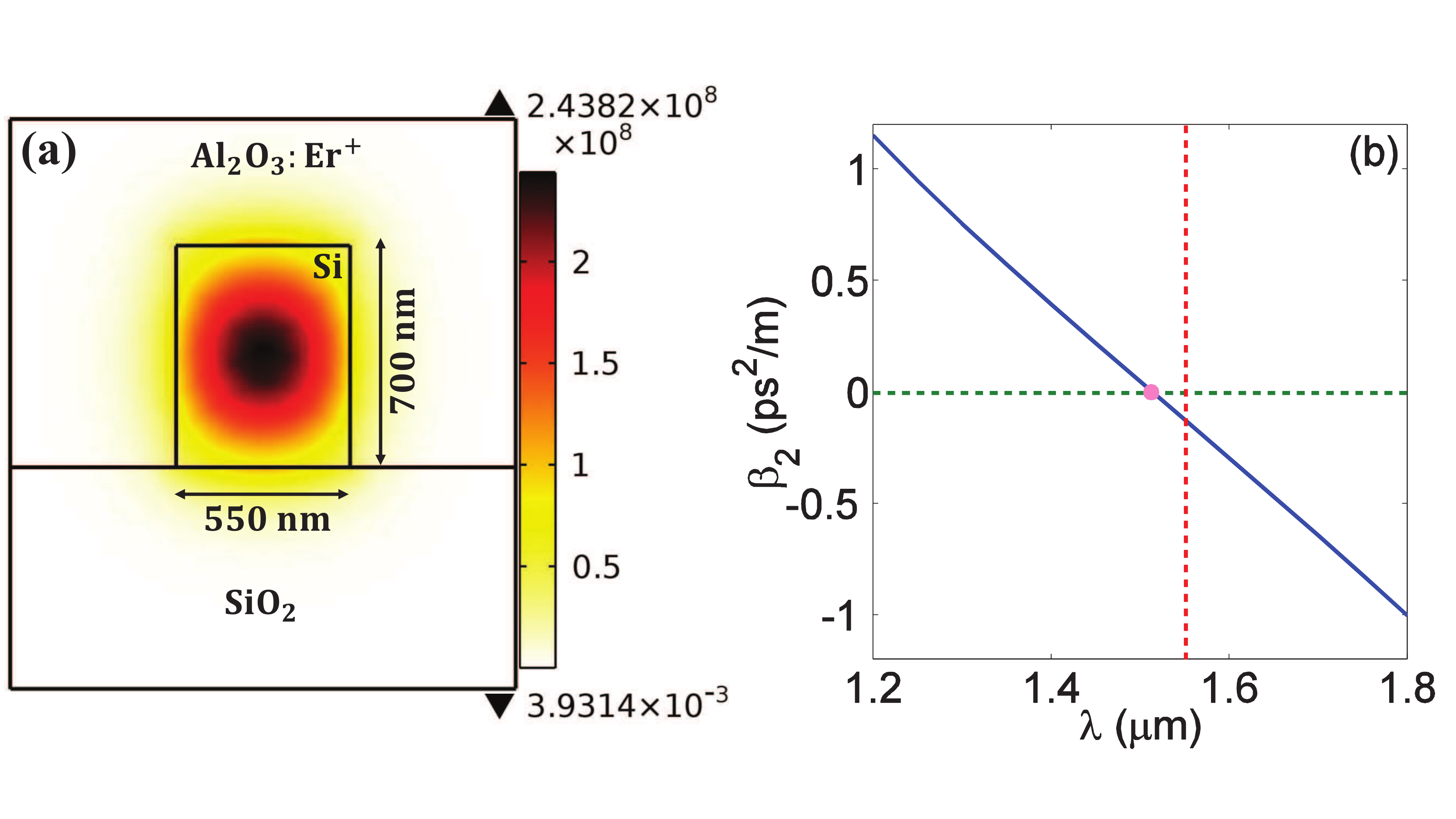,trim=0.00in 0.5in 0.0in 0.75in,clip=true, width=88mm}
 \vspace{0em} 
\caption{(a) The transverse confinement of the quasi-TM mode is shown for $\lambda_0=1.55 \ \mu m$. The width and height of the waveguide are $h=700 \ nm$ and $w=550 \ nm$ respectively. (b) The GVD profile is shown by solid blue line. The location of the input ($\lambda_0=1.55 \ \mu m$) is indicated by vertical dotted red line. The zero dispersion point is shown by the dot. }
\label{structure}
\end{center}
\end{figure}

In Fig.~\ref{structure}(a) the geometry of the waveguide is shown where the fundamental quasi-TM mode is confined in the core. Using commercial COMSOL software we calculate the total dispersion of the waveguide in the wavelength range $1.2\ \mu m$ to $1.8\ \mu m$ (see Fig.~\ref{structure}(b)). The dispersion varies almost linearly with wavelength $\lambda$ and cuts the zero point around $1.5\ \mu m$. At operating wavelength $\beta_2 < 0$, hence the dispersion experienced by the DS is anomalous in nature. The zero dispersion wavelength (ZDW) falls close to the operating wavelength ensuring efficient radiation. The amplitude of the radiation decays significantly when the optical pulse is launched far from the ZDW.

\section{Theory}
\label{theory}
The propagation dynamics of the electromagnetic field with envelope $u(z,t)$  inside the active  waveguide, is modeled by the complex Ginzburg-Landau (GL) equation \cite{Malomed,Roy-M-B},
\begin{align} \label{gl}
i\frac{\partial u}{\partial \xi }-\frac{1}{2}sgn\left( {{\beta }_{2}} \right)\frac{{{\partial }^{2}}u}{\partial {{\tau }^{2}}}-i\left( {{g}_{0}}+{{g}_{2}}\frac{{{\partial }^{2}}}{\partial {{\tau }^{2}}} \right)u+i\alpha u \nonumber \\
-i{{\delta }_{3}}\frac{{{\partial }^{3}}u}{\partial {{\tau }^{3}}}+\left( 1+iK \right){{\left| u \right|}^{2}}u+\left( \frac{i}{2}-\mu  \right){{\phi }_{c}}u=0. 
\end{align}
The FC dynamics due to TPA is realized by an ordinary differential equation $d\phi_c/d\tau=\theta|u|^4-\tau_c \phi_c$ \cite{Lin}. We couple this equation with main GL equation in order to take account the FC effect. We normalize the time $(t)$ and propagation distance $(z)$ as $\tau=tt_0^{-1}$ and $ \xi=zL_D^{-1}$, where $t_0$ and $L_D=t_0^2/|\beta_2 (\omega_0)|$ are initial pulse width and dispersion length respectively. In our calculation we use $t_0 = 25 \ fs$, the time duration of ultrashort pulse. We include the TOD term as perturbation  which is normalized as, $\delta_3=\beta_3/(3!|\beta_2 | t_0)$. We calculate the TOD coefficient using the dispersion profile and find $\delta_3\approx$ 0.23. The field amplitude $(A)$ is normalized as, $A = u\sqrt{P_0}$, where peak power, $P_0=|\beta_2 (\omega_0 )|/(t_0^2 \gamma_R)$. The nonlinear coefficient is defined as, $\gamma_R=k_0 n_2/A_{eff}$ where the Kerr coefficient for Si is, $n_2\approx(4\pm 1.5)\times 10^{-18} \ m^2 W^{-1}$. The normalised TPA coefficient is defined as, $K=\gamma_1/\gamma_R=\beta_{TPA} \lambda_0/(4\pi n_2)$, where for Si  $\beta_{TPA}\approx 8\times 10^{-12} \ m W^{-1}$. The linear loss ($\alpha_l$) is rescaled as, $\alpha = \alpha_l L_D$ which is negligible in the transparency window of Si,  $1\ \mu m <\lambda_0<10\ \mu m$.    The interaction of electromagnetic wave with the semiconductor waveguide leads to the generation of FCs through multiphoton absorption process. In our model we assume the generation of FC is mainly induced by TPA. The density of FCs, $N_c$, generated through TPA is normalized so that $\phi_c=\sigma N_c L_D$. The FC modifies the refractive index through free-carrier dispersion (FCD) and induces loss by absorption. The free-carrier absorption (FCA) cross section of Si at $\lambda_0=1.55\ \mu m$  is $\sigma\approx 1.45\times 10^{-21}\ m^2$\cite{Rong}. The effect of FCA is included in the equation through the parameter $\theta$, normalised as, $\theta=\beta_{TPA} |\beta_2 |\sigma/(2\hbar\omega_0 A_{eff}^2 t_0 \gamma_R^2)$ \cite{Lin-Z-P}. The FCD term is characterised by the parameter $\mu=2\pi k_c/(\sigma \lambda_0)$, where $k_c\approx 1.35\times 10^{-27}\ m^3$ \cite{Dinu}.  In the dynamical equation of FC we include the characteristics recombination time $t_c$ $(\sim ns)$, which is normalized as $\tau_c=t_0/t_c$. The gain  $G$ and gain dispersion coefficient $g_2$ are renormalised as, $g = GL_D$ and $g_2= g(T_2/t_0)^2$, respectively, where $T_2$ being the dephasing time.

\begin{figure}[h!]
\begin{center}
\epsfig{file=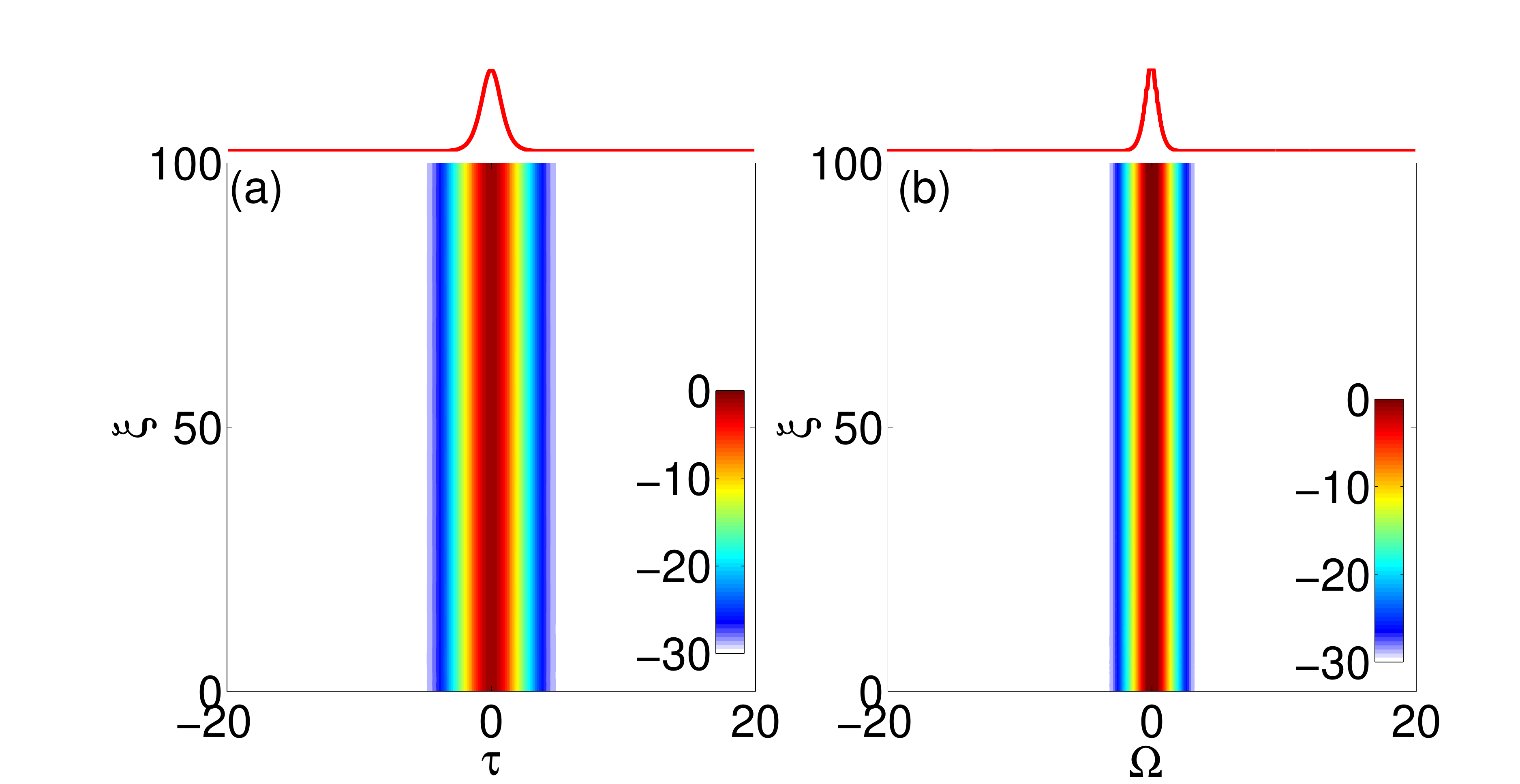,trim=1.1in 0.0in 0.8in 0.65in,clip=true, width=86mm}\vspace{0em}
\caption{(a) Temporal and (b)spectral dynamics of exact solution (Eq.~\eqref{ansatz}) of GL equation in AD regime. Here $\delta_3=0$ , $\phi_c=0$  and $sgn(\beta_2)=-1$ in the GL equation. The overhead line in both panel gives the corresponding output intensity. Here $\Omega=2\pi(\nu-\nu_0 )t_0$. The parameters used in the simulation are: $K=0.01,~g_0=0.01,~g_2=0.01$ and $\alpha=0$.}  \label{exact}
\end{center}
\end{figure}

DSs are formed when a critical dynamical equilibrium is maintained. The parameters of the DS satisfy some unique relationship for stabilizing the formation of the DS. For vanishing  TOD (i.e., $\delta_3=0$) and FC effect (i.e., $\phi_c=0$), the stable solution of the reduced GL equation is in the form \cite{Desurvire},
\begin{equation} \label{ansatz}
u\left( \xi, \tau \right)={{u}_{0}}{{\left[ \text{sech}\left( \eta \tau  \right) \right]}^{\left( 1+ia \right)}}{{e}^{i\text{ }\!\!\Gamma\!\!\text{ }\xi }},
\end{equation}
where the four parameters $u_0,\ \eta,\ a$ and $\Gamma$ are related to the different coefficient of GL equation as, \cite{Roy-M-B,GPA_Appli}:
\begin{subequations} \label{ansatz_part}
\begin{gather}
{{\left| {{u}_{0}} \right|}^{2}}=\frac{({{g}_{0}}-\alpha )}{K}\left[ 1-\frac{\frac{1}{2}sgn\left( {{\beta }_{2}} \right)a+{{g}_{2}}}{\left( {{g}_{2}}\left( {{a}^{2}}-1 \right)-sgn\left( {{\beta }_{2}} \right)a \right)} \right], \\ 
{{\eta }^{2}}=\frac{({{g}_{0}}-\alpha )}{\left( {{g}_{2}}\left( {{a}^{2}}-1 \right)-sgn\left( {{\beta }_{2}} \right)a \right)}, \\ 
\Gamma=\frac{{{\eta }^{2}}}{2}\left[ sgn\left( {{\beta }_{2}} \right)\left( {{a}^{2}}-1 \right)+4a{{g}_{2}} \right], \\
a=\frac{H-\sqrt{{{H}^{2}}+2{{\delta }^{2}}}}{\delta }.
\end{gather}
\end{subequations}
Here, $H=-[(3/2)sgn(\beta_2 )+3g_2 K]$ and $\delta=-[2g_2-sgn(\beta_2)K]$. The  values of $u_0,\ \eta,\ a$ and $\Gamma$ are uniquely determined by the physical parameters : $g,\ g_2,\ K$ and $\alpha$. We use Eq.\eqref{ansatz} as input and numerically solve the unperturbed (i.e., $\delta_3=0$ and $\phi_c=0$) GL equation.  The numerical solution is shown in Fig.~\ref{exact}, where we see that the structure of the DS is robust and propagating without any distortion. This justifies the validity of the derived relations in Eq. \eqref{ansatz_part}. In the following section we solve  the extended GL equation containing TOD and show how DWs are generated.

\section{Formation of dw}
\label{dw}
The TOD is responsible for the generation of temporal side lobes during the propagation of a soliton \cite{GPA}. The non-zero TOD also acts as a higher order modification in the group delay of the pulse. New term appears due to TOD when we expand the propagation constant $\beta(\omega)$  in Taylor series around DS frequency. This new term containing TOD coefficient, leads to a real solution of the PM equation which ensures the energy transfer from soliton to a linear wave \cite{Wai}. During the supercontinuum generation process, the blue component of the spectra are mainly generated because of this phase-matched radiation \cite{Yin-L-GP,Dudley-T}. The implication of this resonant radiation and its control are studied extensively in the context of perturbed soliton dynamics \cite{Roy,GPA}. However this problem is less addressed when the system is non-integrable and governed by some energy balance condition. Hence it is interesting to explore the generation and behaviour of such radiation when, instead of Kerr soliton, the system is excited by a DS. In our study, firstly, we confirm that, like Kerr soliton the DS also radiates when it is perturbed by higher order dispersion. A fundamental radiation in the form of DW appears for a short distance of propagation. The nature of the radiation becomes polychromatic when the DS travels a larger distance. We derive a general PM equation in analytic form which satisfactorily predicts the radiation frequency (or frequencies) in both the cases.

\begin{figure}[h!]
\begin{center}
\epsfig{file=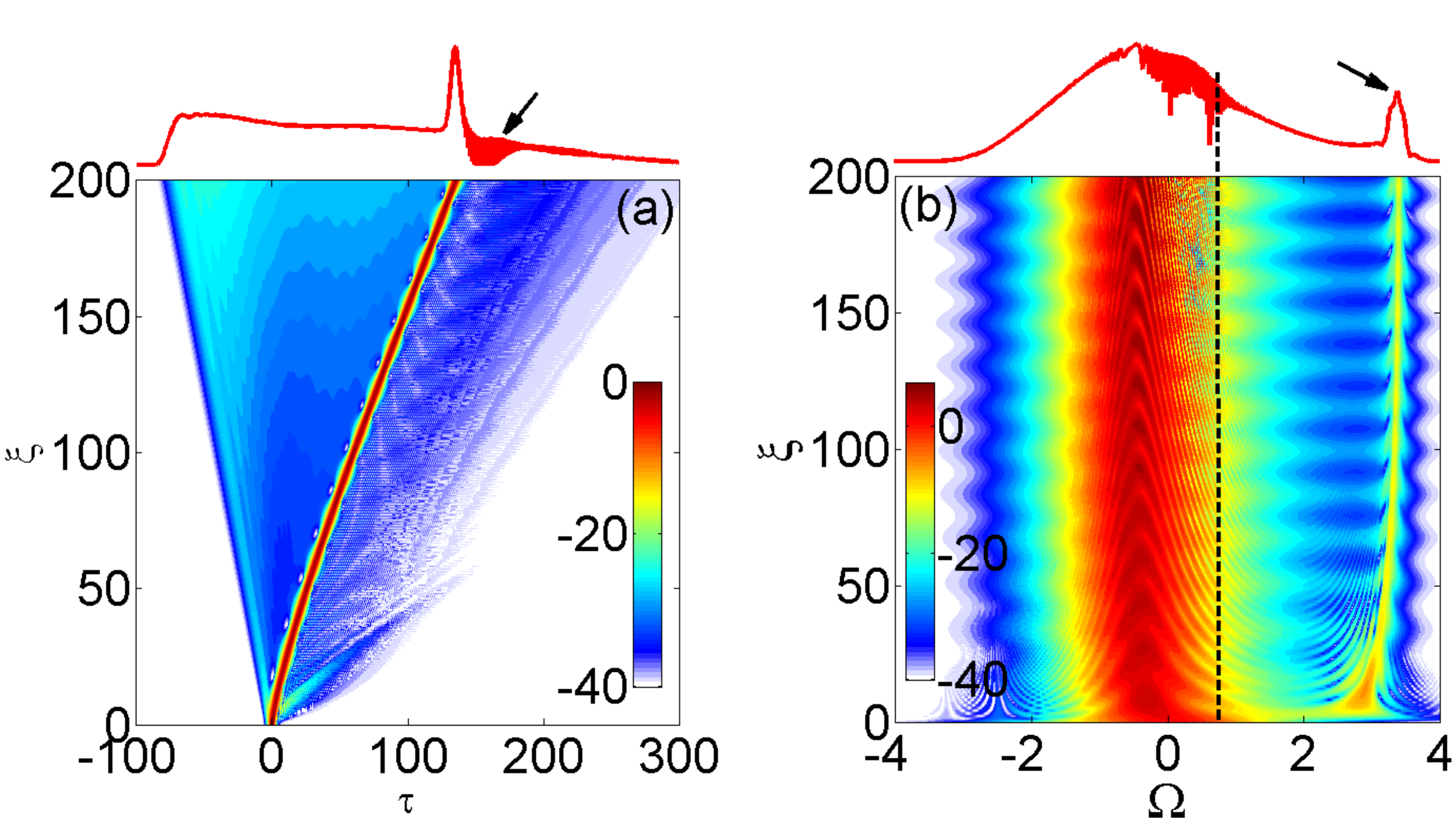,trim=0.0in 0.0in 0.0in 0.4in,clip=true, width=85mm}
\vspace{0em}
\epsfig{file=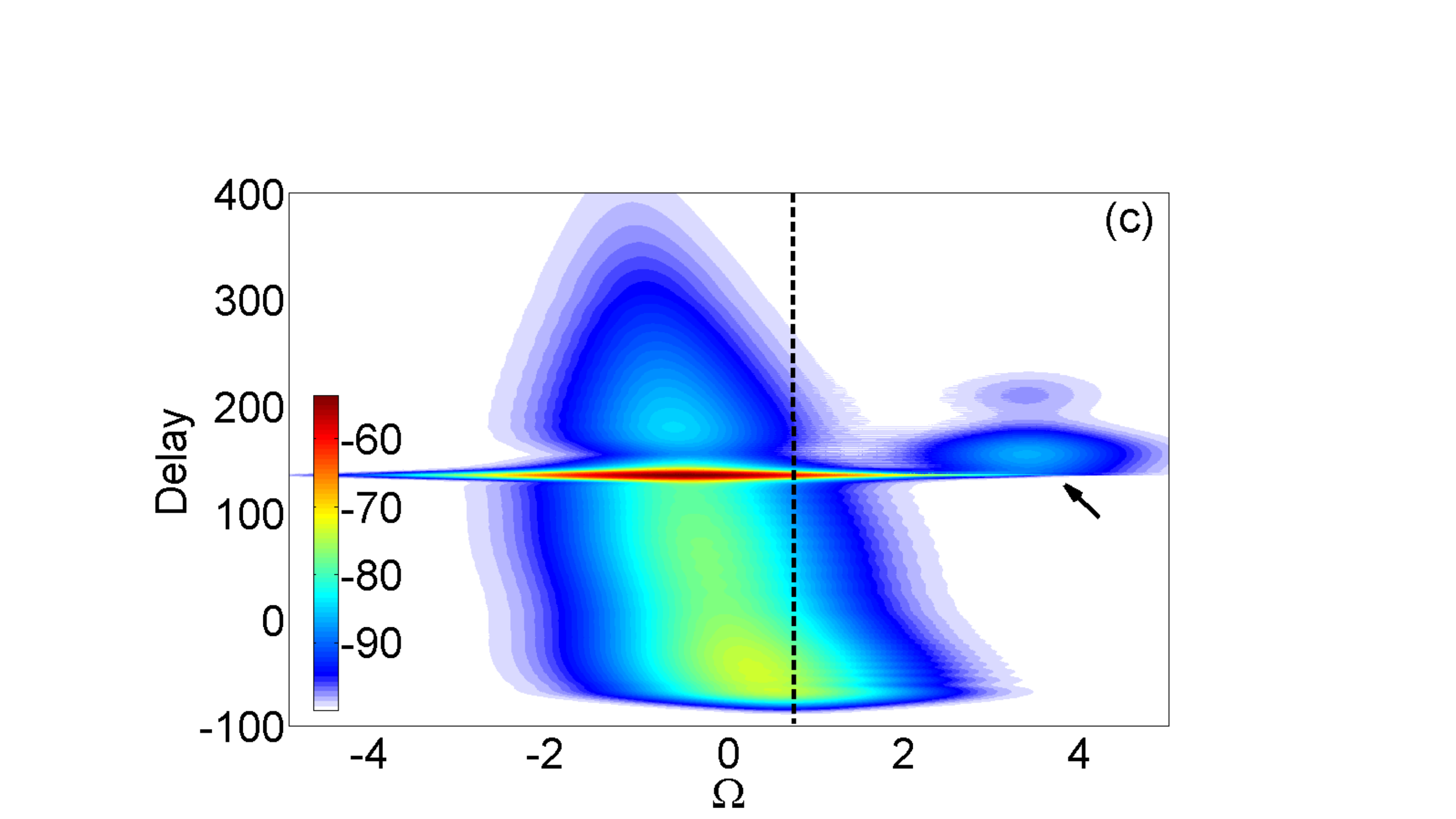,trim=1.4in 0in 2.5in 1.4in,clip=true, width=65mm}
\vspace{0em}
\caption{ (a) Temporal and (b) spectral evolution of DS under the action of TOD $(\delta_3=0.23)$. The overhead line gives corresponding output intensity plot. (c) Spectrogram at $\xi=200$ for $\delta_3=0.23$. The arrows indicate the signature of DW in all figures. The vertical black dotted line in (b) and (c) represents the position of zero GVD position $\Omega_{ZGVD}\approx 0.74$. The other parameters are taken to be same as Fig.~\ref{exact}.}  \label{DWgen}
\end{center}
\end{figure}
 
\begin{figure}[h!]
\begin{center}
\epsfig{file=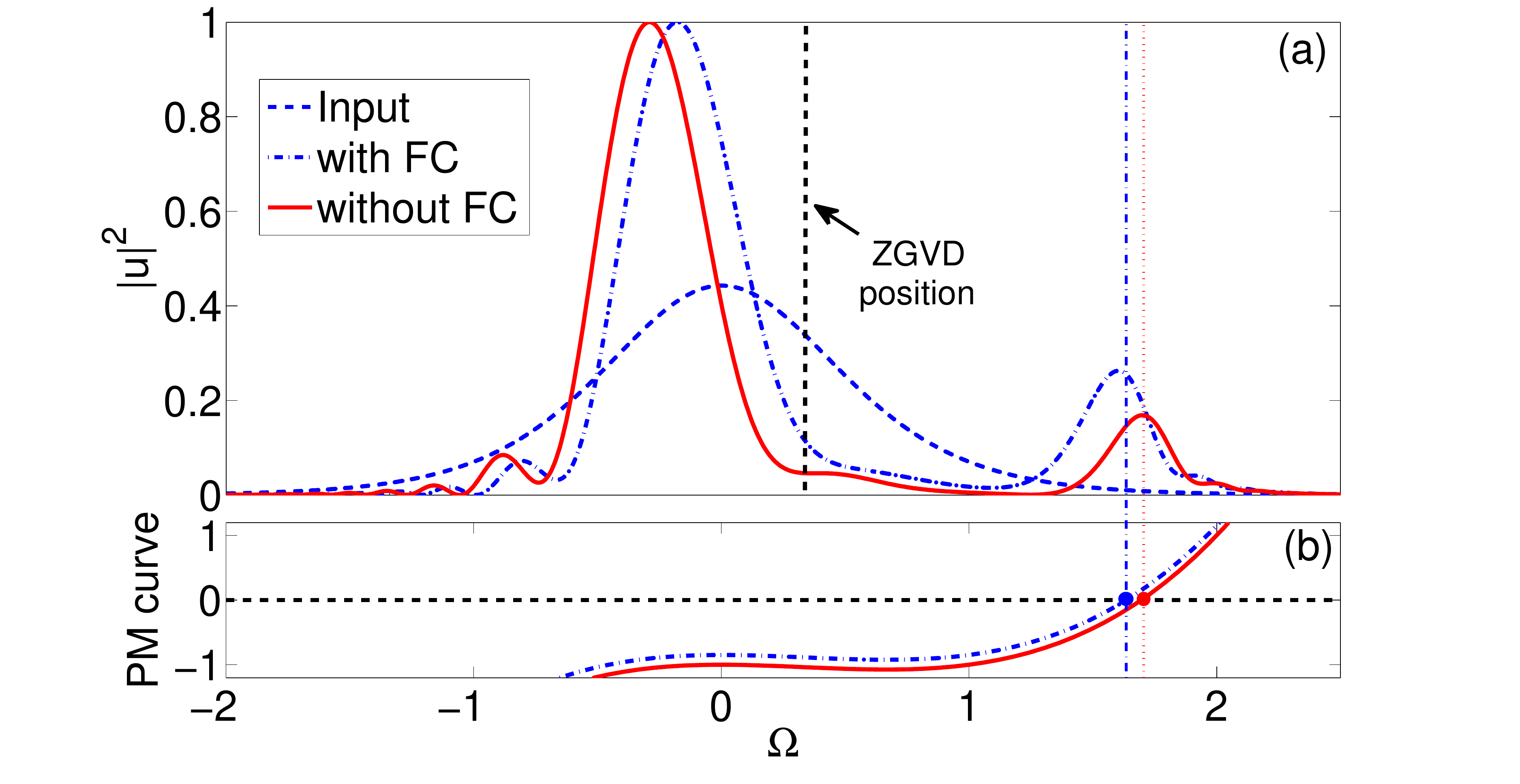,trim=1.0in 0.25in 1.5in 0.05in,clip=true, width=87mm}
\vspace{0em}
\caption{(a) Output spectrum of the DS at $\xi= 10$ in presence (dot-dashed blue) and absence (solid red) of FC.  The vertical black dotted line indicates the location of zero GVD point. (b) Graphical solution of Eq.~\eqref{pm_FC} in presence (dot-dashed blue) and in absence (solid red) of FC. The vertical dashed lines are the solutions which indicate the location of DWs. The parameters, that we have taken for the simulations are $\delta_3=0.5$, $K=0.01$, $g_0=0.01$, $g_2=0.01$, $\alpha=0$, $\theta=0.04$.}  \label{FC}
\end{center}
\end{figure}

The numerical solution of GL equation containing TOD and FC term gives us the complete dynamical picture of the propagating DS. In Fig.~\ref{DWgen} we try to capture the temporal and spectral evolution of the perturbed DS. Here one can see that, DS is influenced by TOD and DWs are generated when the spectrum overlaps the zero dispersion boundary shown by the vertical dotted lines in the figures. Using the density plot in Fig.~\ref{DWgen}(a) and Fig.~\ref{DWgen}(b) we show the temporal and spectral evolution of a DS. To identify the spectral counterpart of a given temporal pulse simultaneously, we adopt cross-correlation frequency resolved optical grating (XFROG) spectrogram technique. By definition, the XFROG is mathematically represented as the convolution $s(\tau,\omega,\xi)=|\int_{-\infty}^{\infty} u(\xi,\tau')u_{ref}(\tau-\tau')\exp(i\omega\tau')d\tau'|^2$ , where $u_{ref}$ is the reference window function normally taken as an input \cite{GPA}. 
 In Fig.~\ref{DWgen}(c) we depict the XFROG of the DS at $\xi=200$, where formation of the DW is evident and indicated by an arrow. The stable DS oscillates with certain breathing period ($\xi_0$) when, the contribution of TOD is taken into account in the simulation. We numerically examine that, $\xi_0$ is mainly a function of $K$ and $g_0$. From Fig.~\ref{DWgen} (b), it is clear that  multiple radiations are created across zero GVD point at each temporal compression cycle of DS. The temporal and spectral evolution of DS is shown in Fig.~\ref{DWgen}(a) and (b), respectively. Multiple DWs manifest interference, and comb-like radiation band may appear. The resonance with the zero temporal harmonic is referred to the Cherenkov radiation \cite{Akhmediev-K} whereas, the secondary harmonics generate additional radiation peaks \cite{Kodama,Confort,Driben,Yulin}. Before going to the detailed study of multiple DW radiation, it is important to investigate the evolution of the fundamental DW. In the semiconductor waveguides like Si, FCs are generated as a consequence of multi-photon absorption and are dynamic in nature \cite{Rukhlenko}. To grasp the role of FC on DW generation, we numerically solve the coupled GL equation by including FC effect. The numerical solutions with (dotted line) and without (solid line) FC are compared in Fig.~\ref{FC}(a). From  the figure, it is observed that, in presence of FC, the main DS shifts towards higher frequency side, whereas the DS shifts towards lower frequency side from its original position (the case when FC is not considered). Even though the frequency shift of DW due to FC is relatively small, it is possible to capture this effect analytically. The additional phase induced by the FC modifies the PM equation and is responsible for the small shift of the resonance frequency. The rate equation for FC dynamics has an exact analytical solution  when recombination time is large ($\tau_c\rightarrow0$) \cite{SR-SKB-GPA}. Exploiting the solution of $\phi_c$,   it is possible to calculate the average phase induced by FCs.  After propagating a distance of $\xi$, DS acquire an addition phase  $\approx -(2/3)\mu \theta u_0^4 \xi /\eta$ due to FCs. This additional phase should be included during the calculation of PM condition. For an oscillating DS,  the resonance condition (in normalized unit) takes the general form:  

\begin{equation} \label{pm_FC}
\frac{sgn\left( {{\beta }_{2}} \right)}{2}{{\Omega}^{2}}+{{\delta }_{3}}{{\Omega}^{3}}-{{\tau }_{g}}\Omega=\frac{2\pi n}{{{\xi }_{0}}}+2\Gamma - \frac{2}{3}\mu \theta \frac{u_0^4}{\eta}.
\end{equation}
Here, $\Omega=2\pi(\nu_d-\nu_s )t_0$ is the detuning frequency, $\xi_0$ is the breathing period and $n$ is an integer number $(n=0,1,2,3,...)$. The group delay mismatch of DS is defined by $\tau_g$. For small distance, when the breathing of DS is yet to begin and the group velocity mismatch between DS and DW is very small, we can approximate $n=0$ and $\tau_g=0$. In Fig.~\ref{FC}(b), we plot the PM curve whose zero crossing point gives the solution. The PM solutions accurately predict the spectral locations of the DW in presence (dot-dashed blue line) and in absence (solid red line) of FC.  

\begin{figure}[h!]
\begin{center}
\epsfig{file=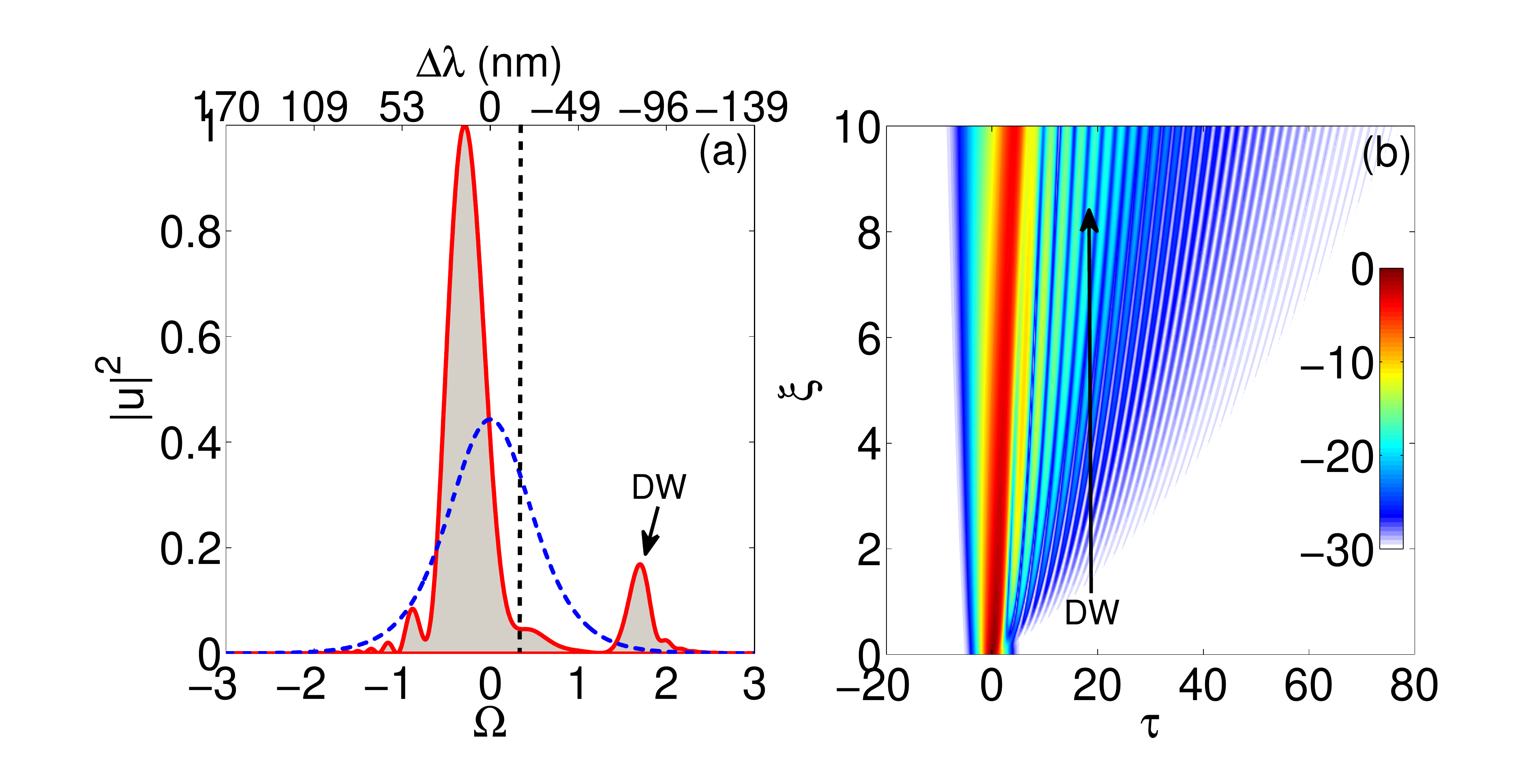,trim=0.9in 0.2in 0.8in 0.4in,clip=true, width=87mm}\vspace{0em}
\epsfig{file=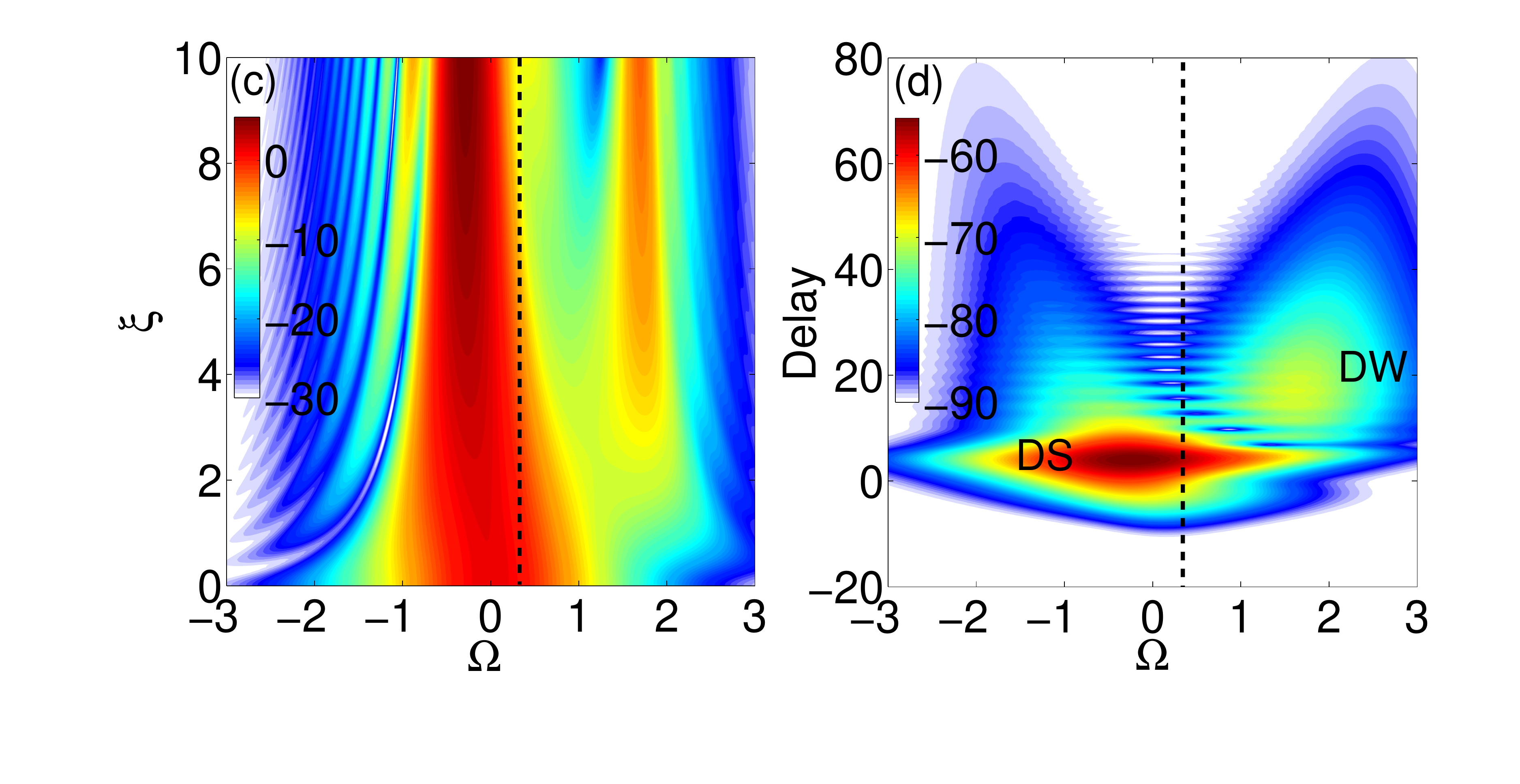,trim=0.9in 0.85in 0.8in 0.4in,clip=true, width=87mm}\vspace{0em}
\caption{(a)  Output spectrum of the DS at $\xi= 10$. The input is indicated by blue dotted line.  The hump due to the DW in normal dispersion regime is shown by the arrow. (b) Time and (c) spectral evolution of the DS  where the generation of DW is evident at $\Omega\approx 1.75$ . (d) The corresponding spectrogram at $\xi= 10$. The parameters, that we have considered for the simulations  are $K=0.01$, $g_0=0.01$, $g_2=0.01$, $\xi=10$, $\alpha=0$ and $\delta_3=0.5$.}  \label{detuned-in-out}
\end{center}
\end{figure}

In order to ensure the validity of the PM expression (as shown in Eq.\eqref{pm_FC}) for a wide range of TOD parameters, we apply it for fundamental DW (i.e., $n=0$) called Cherenkov radiation. For the fundamental DW the PM equation reduces to,
\begin{equation} \label{pm4}
\frac{sgn\left( {{\beta }_{2}} \right)}{2}{{\Omega}^{2}}+{{\delta }_{3}}{{\Omega}^{3}}=2\Gamma - \frac{2}{3}\mu \theta \frac{u_0^4}{\eta}.
\end{equation}

The fundamental phase-matched radiation appears within a very short distance where the relative delay between DS and linear DW can be neglected (i.e., $\tau_g=0$).  Eq.~\eqref{pm4} can be solved  graphically. Note, by putting $\Gamma=0$ (for conventional soliton) and  $\theta=0$ (no FC) , the Eq.~\eqref{pm4} can be reduced to well-known PM equation which has an exact solution of the form \cite{GPA},
\begin{equation} \label{pm5}
{{\Omega}_{0}}=-\frac{sgn\left( {{\beta }_{2}} \right)}{2{{\delta }_{3}}}.
\end{equation}

\begin{figure}[h]
\begin{center}
\epsfig{file=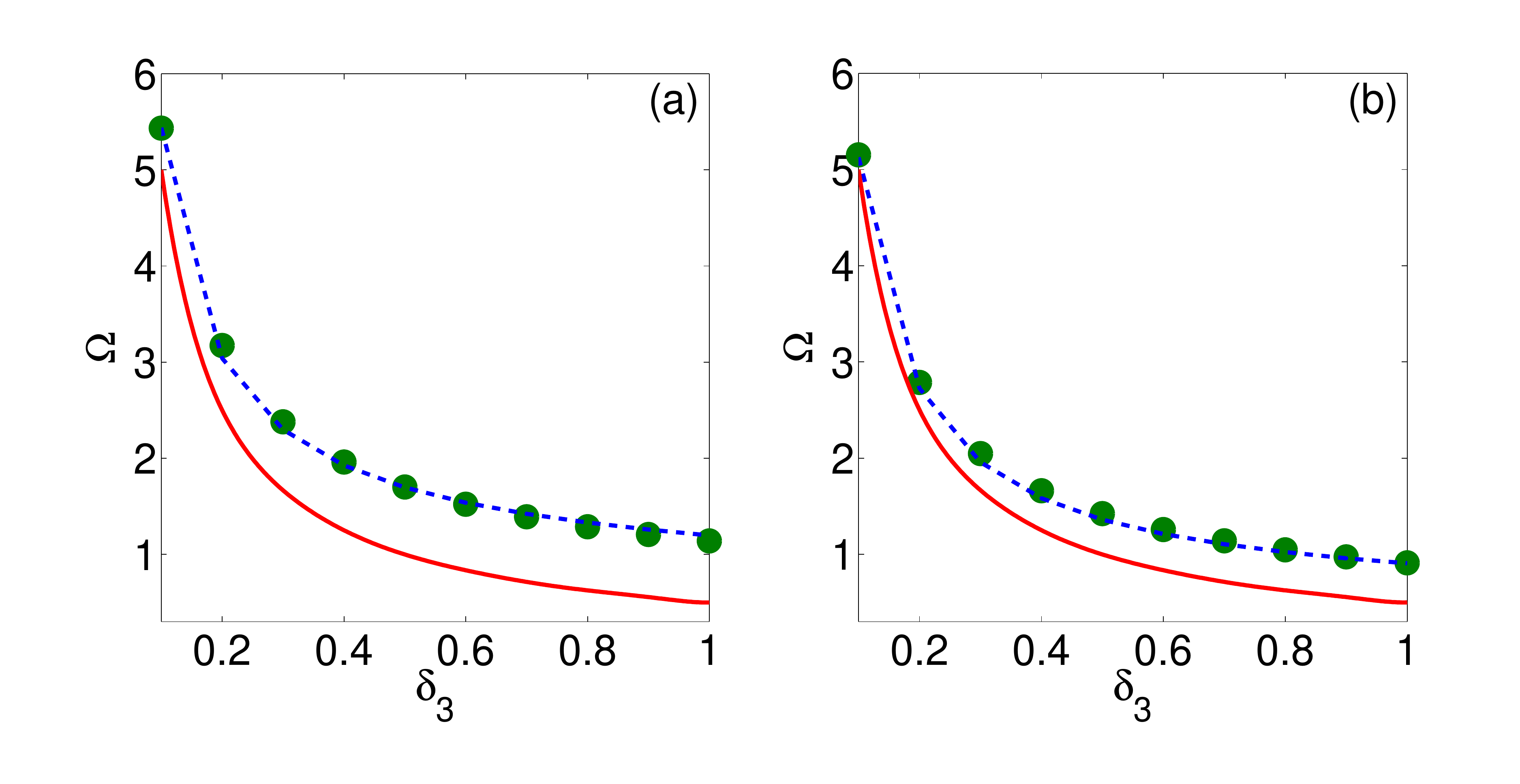,trim=0.45in 0.6in 0.7in 0.5in,clip=true, width=88mm}\vspace{0em}
\epsfig{file=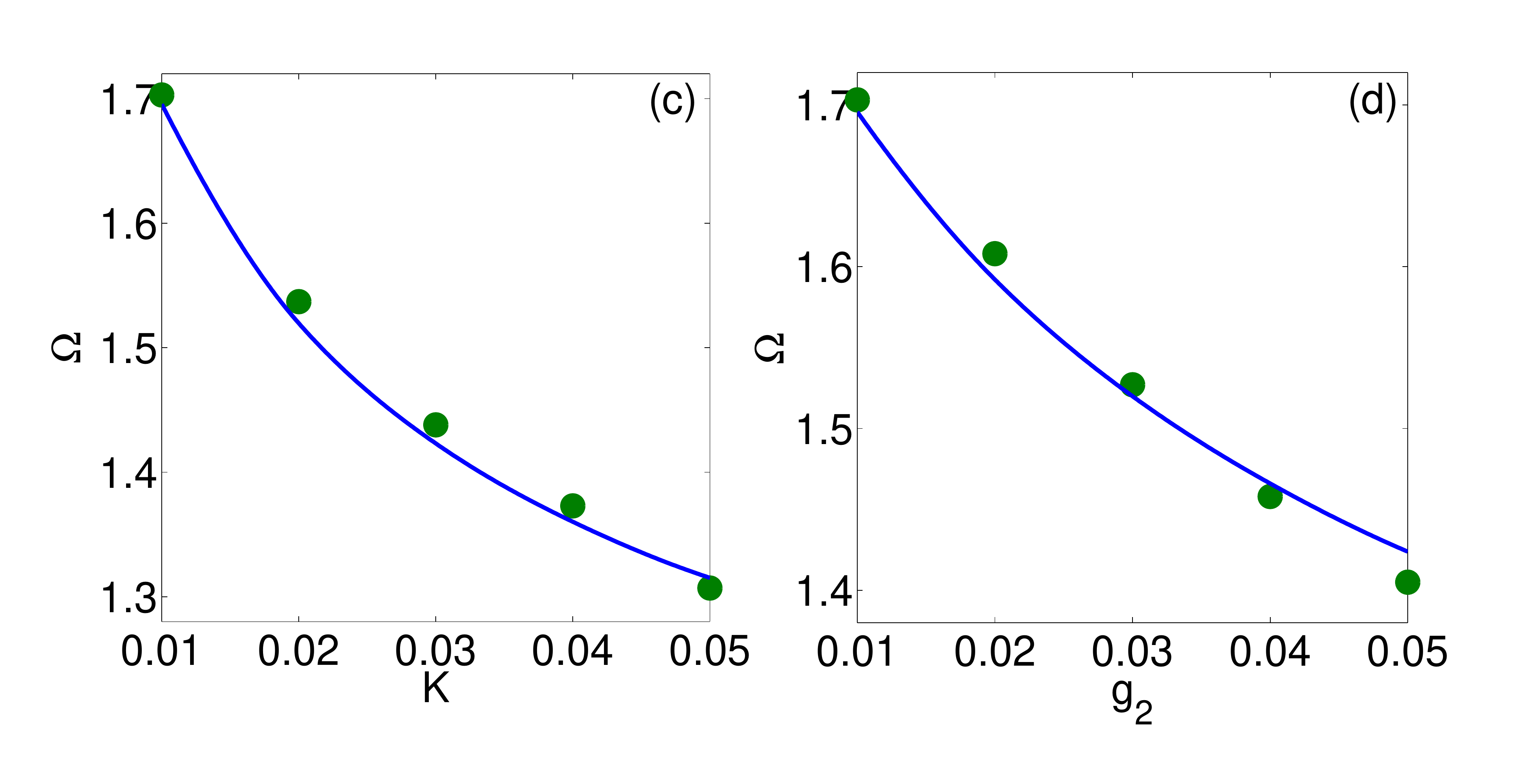,trim=0.45in 0.6in 0.7in 0.5in,clip=true, width=88mm}\vspace{0em}
\caption{(a) Variation of radiation frequency with $\delta_3$ (a) for $K=0.01$  and (b) $K=0.04$ respectively. The dotted curve represents the solution of Eq.~\eqref{pm4} which agrees well with numerical results (solid dots). The solid curve represnts the frequency location predicted by the existing theory (Eq.~\eqref{pm5}) which disagrees  with the numerical results. The variation of radiation frequency with (c) $K$ and (d) $g_2$ is shown when TOD parameter is kept fixed ($\delta_3=0.5).$ The proposed theory (solid lines) agrees well numerical results (dots). For simulation the parameters we used are: $\xi=10$, $g_0=0.01$, $g_2=0.01$ and $\alpha=0$.}  \label{detuned}
\end{center}
\end{figure}

Exploiting split-step Fourier technique \cite{GPA}, we numerically solve the GL equation for non-zero TOD coefficient and show all the results in Fig.~\ref{detuned-in-out}. A distinct peak (around $\Omega\approx$ 1.75) is observed in Fig.~\ref{detuned-in-out}(a) indicating resonance frequency as a result of energy transfer. In temporal evolution (see Fig.~\ref{detuned-in-out}(b)), the linear DW is prominent as a side lobe which is moving with slightly different velocity. The spectral evolution and XFROG are also shown in Fig.~\ref{detuned-in-out}(c) and Fig.~\ref{detuned-in-out}(d) respectively.
At relatively small distance of propagation ($\xi=10$), DS excites the fundamental single radiation which is clearly visible both in Fig.~\ref{detuned-in-out}(c) and Fig.~\ref{detuned-in-out}(d). In Fig.~\ref{detuned}(a) and (b) we plot the fundamental resonance frequency as a function of $\delta_3$ for two distinct TPA coefficients ($K=0.01,0.04$). In the figures, dots represent the numerical data, whereas the dashed line corresponds to the solution obtained from Eq. \eqref{pm4}. The analytical result agrees well with simulation data. The additional solid lines in  Fig.~\ref{detuned}(a) and (b) indicate the radiation frequency derived from Eq.~\eqref{pm5} which is the standard solution for the radiation emitted by Kerr soliton. The solid lines clearly misfit the numerical data (dots). So we can infer that, the conventional PM solution (Eq.~\eqref{pm5}) is incompetent to predict the spectral location of DS-mediated radiation. The modified PM expression also unfold that, for a fixed $\delta_3$ the DS-mediated radiation can be tailored by the TPA parameter $K$ and gain dispersion $g_2$.  This unique feature is absent in conventional Kerr soliton where $\delta_3$ is the only parameter that controls the radiation frequency. In Fig.~\ref{detuned}(c) and (d) we show the variation of the detuned frequency as a function of $K$ and $g_2$ when TOD parameter is fixed ($\delta_3=0.5$). The theoretical predictions are shown by the solid lines whereas dots represent numerical data. 

\begin{figure}[h!]
\begin{center}
\epsfig{file=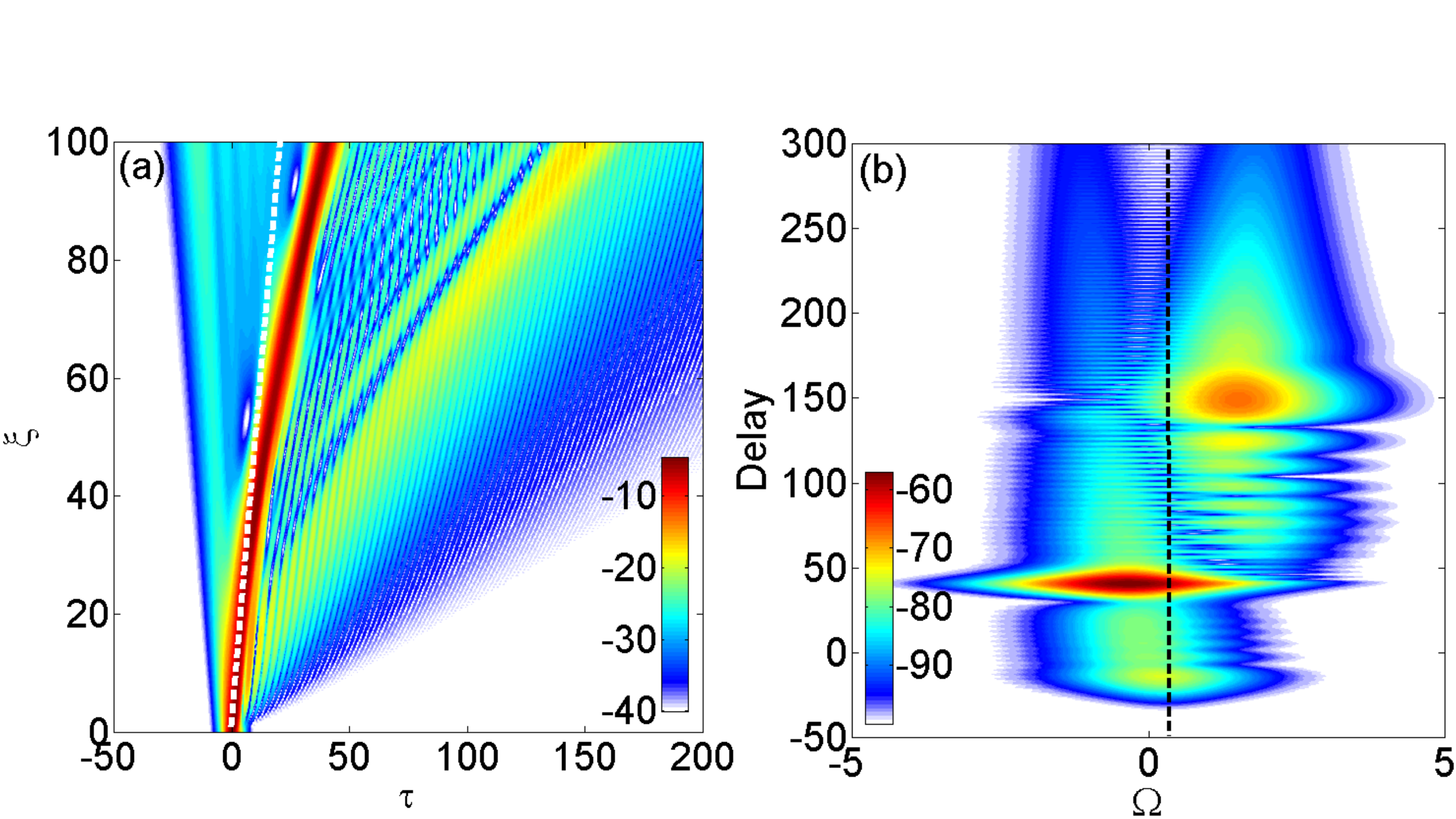,trim=-0.2in 0in 0in 1.1in,clip=true, width=87mm}
\vspace{0em}
\epsfig{file=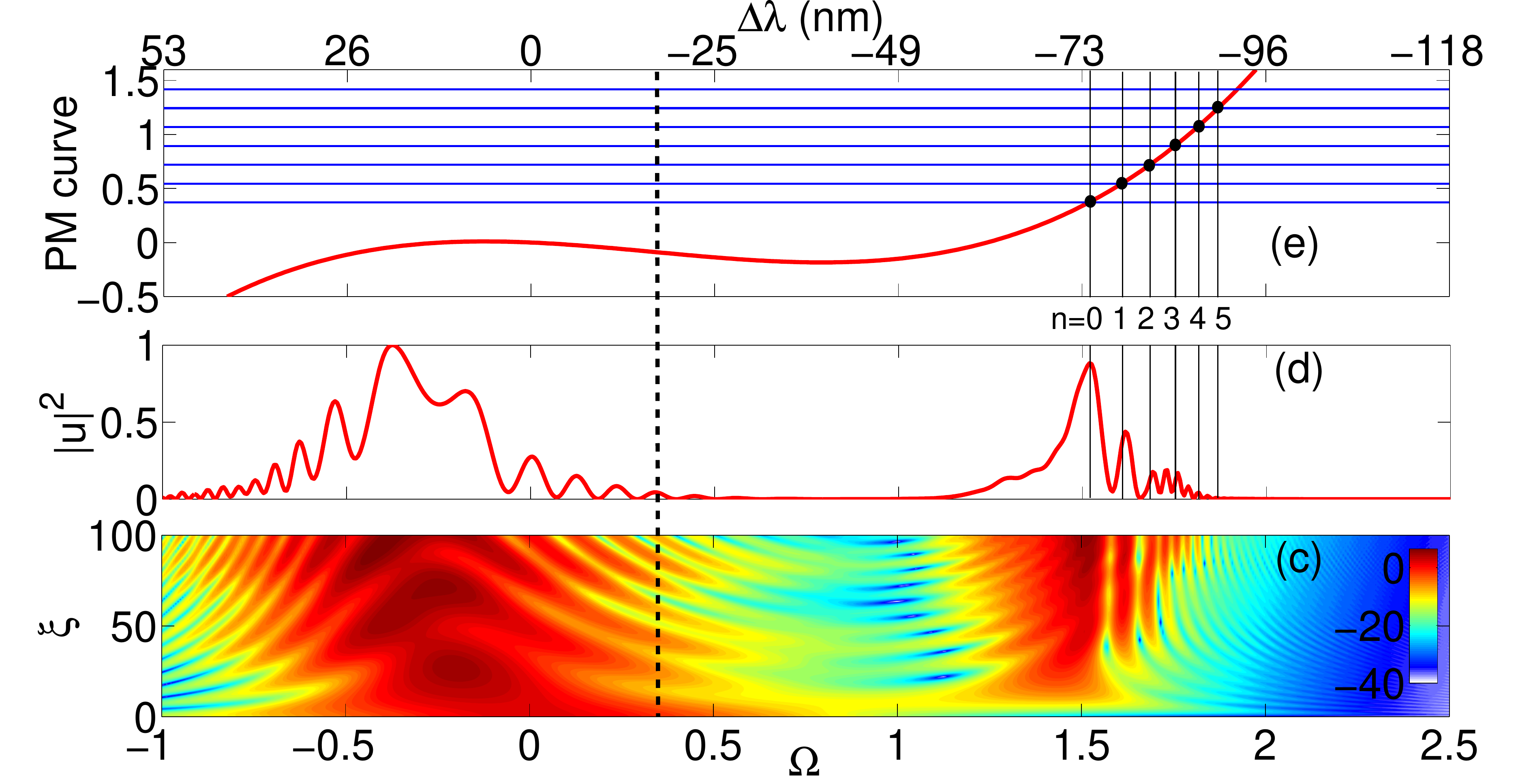,trim=0.35in 0.05in 0.38in 0.0in,clip=true, width=88mm}
\vspace{0em}
\caption{The evolution of DS in  (a) time and (c) frequency domain,  respectively. The white dotted line in (a) shows the relative group delay mismatch which is  $\tau_g\approx 0.15$. (b) The spectrogram at $\xi=100$. In the simulation we used $\delta_3=0.5$, $K=0.04$, $g_0=0.01$, $g_2=0.001$, $\alpha=0$, $\xi_0=36$, $\Gamma=0.1851$.  (d) Output spectrum shows  multiple radiation picks. (e) Graphical solution of Eq.~\eqref{pm_FC} indicating spectral positions of multiple resonance radiations. The vertical dotted line indicates the position of zero GVD position. }  \label{multipleDW}
\end{center}
\end{figure}

DS exhibits multiple radiations when it is allowed to propagate through a large distance under low gain dispersion condition. In such case, the DS evolves periodically and emits every time when the corresponding spectrum crosses the zero dispersion boundary. In  Fig.~\ref{multipleDW} we capture the dynamics of a DS when it is allowed to travel over a distance $\xi=100$ with gain dispersion coefficient $g_2=0.001$. The temporal dynamics (Fig.~\ref{multipleDW}(a)) shows that the fundamental DW is emitted from the DS as a result of energy transfer followed by some secondary radiations. These  secondary radiations exhibit multiple peaks in the spectral domain as shown in Fig.~\ref{multipleDW}(d). In the spectrogram (Fig.~\ref{multipleDW}(b)) we capture the formation of multiple DW at a distance $\xi=100$. The dynamic evolution of these multiple radiations are also demonstrated in the density plot shown in Fig.~\ref{multipleDW}(c). The graphical solution of Eq.~\eqref{pm_FC} predicts the spectral location of the individual radiation convincingly as  shown in Fig.~\ref{multipleDW}(e). The precise matching of analytical and numerical results confirms the validity of our proposed PM relation for DS which was not explored before.

\section{Conclusion}
In conclusion, we have shown that, stable DS radiates DWs when it is  perturbed by TOD. An active Si-based waveguide is proposed that can support stable DS generation. The dispersion property of the waveguide is engineered in such a way that the spectral gap between operating and zero dispersion frequency is small, which is an important condition to excite strong resonant radiation. TPA and gain dispersion are found to be the two additional parameters that can also influence the radiation frequency. This feature is unique for DS-mediated radiation. By including the effect of FCs, we derived a general PM equation which predicts the spectral location of the radiation. When the complete numerical simulation is performed for a long distance, we find DS exhibits multiple radiations. We examine that, under the small perturbation of TOD, DS evolves periodically and radiates every time when its spectral components overlap zero dispersion frequency. The graphical solution of the general PM equation can also locate the frequencies of these multiple radiations precisely.

\section*{Acknowledgement}
This work is supported by SRIC, Indian Institute of Technology Kharagpur, under the project ISIRD. A.S. acknowledges MHRD, India for a research fellowship.

\ifCLASSOPTIONcaptionsoff
  \newpage
\fi

\end{document}